\documentclass[12pt,preprint]{aastex}

\slugcomment{} 
\shorttitle{The Luminosity of VY CMa}
\shortauthors{Humphreys}  

\begin{document}


\title{VY Canis Majoris: The Astrophysical Basis of Its Luminosity}  


\author{Roberta M. Humphreys}
\affil{School of Physics and Astronomy, University of Minnesota,
    55455}



\begin{abstract}
The luminosity of the famous red supergiant VY CMa ( $L \sim  4 - 5 \times 10^{5} L_{\odot}$) is well-determined from its spectral energy distribution 
and distance, and places it near the empirical upper luminosity limit 
for cool hypergiants. In contrast, its surface temperature
is fundamentally ill-defined. Both contradict a recent paper by Massey, 
Levesque and Plez (2006). Implications for its location on the HR Diagram and 
its apparent size are discussed.  
\end{abstract}





\section{Introduction}
Massey, Levesque \& Plez (2006) suggest that the 
famous red supergiant VY CMa is nearly a factor of 10 less luminous 
than has previously been stated 
by several authors. They assert that its high luminosity and 
other ``extreme'' properties such as its inferred large size were based 
on an adopted effective temperature that was too low (e.g. $\sim$ 2800$\arcdeg$, Le Sidaner  
\&  LeBetre (1996)).

Massey et al fit  recent optical spectrophotometry of VY CMa 
with  MARCS
model atmospheres and derive a much warmer effective temperature.  
Combining  its apparent visual magnitude, an adopted 
interstellar extinction, and a temperature dependent bolometric correction,  
they derive a luminosity $L  \sim 6  \times 10^{4} L_{\odot}$ 
instead of the usually quoted $L  \sim 5 \times 10^{5} L_{\odot}$. 
However this classical  approach ignores one 
of VY CMa's  distinguishing 
characteristics, its spectral energy distribution and large excess 
radiation in the  infrared. In this brief paper I  discuss 
VY CMa's energy distribution, its resulting luminosity and location on the HR Diagram.

\section{ The  Energy Distribution and Luminosity} 

Figure 1 shows VY CMa's spectral energy distribution from 0.5 to 100 $\mu$m. 
The data are based on recent photometry in Tables 3 and 4 in Smith et al (2001) plus   
the IRAS fluxes from 25 to 100 $\mu$m. Although VY CMa is an irregular variable
in the visual,  its long wavelength fluxes have shown only small variability.
Most of the star's radiation is reprocessed by the dust in its extensive 
circumstellar ejecta. I  therefore used the total fluxes integrated over 
the entire nebula for 
Figure 1. Its  energy distribution rises rapidly in the infrared and has a broad 
maximum between 5 and 10 $\mu$m.  Integrating the apparent energy distribution 
yields a luminosity of 
 $L  = 4.3 \times 10^{5} L_{\odot}$ 
 at VY CMa's distance of 1.5 kpc (Herbig 1972, Lada \&  Reid 1978, Marvel 1997, 
 the same distance used by Massey et al.) 
If corrected for {\it interstellar} extinction at visual and red wavelengths, the luminosity 
would  increase by a few 
 percent.\footnote{Essentially the same luminosity is obtained whether or not a correction 
 for interstellar extinction is applied because most of the flux is escaping at $\sim$ 10$\mu$m. 
 Furthermore, an A$_{v}$ of 3.2 mag (Massey et al)  implies that at least 2 mag of 
 more of circumstellar 
 extinction is required in the visual to equal  the flux emitted at 10$\mu$m. The wavelength 
 dependence of the CS extinction correction, however, is not known.}  

The  standard ``textbook'' approach, relying only on visual photometry and
spectroscopy and  an  assumed temperature, is not valid for stars 
with sufficient circumstellar dust to reradiate their visual and red flux 
in the thermal infrared. In some cases, the radiating dust also dominates  the  
radiation between   1 $\mu$m and 5 $\mu$m and contributes significant circumstellar 
extinction  at visual, red and near-infrared wavelengths.
Other well-studied examples in 
our galaxy are VX Sgr, S Per, and NML Cyg. Like VY CMa, all three are strong
maser sources and NML Cyg is optically obscured. See Schuster, Humphreys \&  Marengo (2006) 
for recent images of these stars. 

In summary, the luminosity proposed for VY CMa by Massey et al is far less than what is
actually observed, and there is little doubt that it is near the empirical upper luminosity 
limit for the cool hypergiants (Humphreys \&  Davidson 1979, 1994).

\section{ Discussion -- VY CMa on the HR Diagram}

There is general consensus that VY CMa is a red supergiant. Its high luminosity is
well-determined from its spectral energy distribution and distance.
Further consideration of its exact  position on the HR Diagram  depends on the assumed 
surface temperature. Previously published spectral types for VY CMa in the past 30 years
or so have been mostly in the M4-M5 range; however, Massey et al suggest that 
VY CMa's apparent spectral type is more likely $\sim$ M2.5 based on the 
MARCS model atmosphere fit to their spectrum. 
Interestingly, though, the blue TiO bands  in their published spectrum 
(Figure 2 in Massey et al) are more like  
their M4-type  reference spectrum than the M2-type spectrum they show. This author's 
numerous spectra of VY CMa obtained over many years have all been in the M4-M5 range. 
Adopting
this spectral type with the temperature scale proposed by Levesque, et al (2005) 
gives T$_{eff}$ $\sim$ 3450--3535$\arcdeg$, while an older scale (used in Humphreys \&  McElroy
1984 from Flower 1977) yields T$_{eff}$ $\sim$ 3200$\arcdeg$ for an M4-M5 star.

However, one should be cautious in the case of VY CMa; 
{\it we are not observing either its 
photosphere or its surface directly.} It has been known for some time that VY CMa's absorption
spectrum is significantly redshifted with respect to its systemic velocity
(Humphreys 1975, Wallerstein 1977) due to scattering by dust (Herbig 1970,
Kwok 1976, Van Blerkom \&  Van Blerkom 1978).
Indeed, most of VY CMa's visual-red radiation originates by reflection and scattering
by the dust grains at 100 AU from the star, the dust formation radius. Only a few percent 
of the  radiation actually escapes through the dust shell, which is very likely
inhomogeneous,   
implying optical depths of 4 to 5 at $\sim$ 7000$\AA$ in its wind (Humphreys et al 2005).
If the wind is opaque, then R$_{ph}$ where the photons arise, could be larger than the
true stellar radius, and the underlying star possibly somewhat
warmer. 

Massey et al also suggest that with previous temperature estimates, VY CMa would
violate the Hayashi limit.   
But the cause of the apparent conflict with the Hayashi limit is the assumed temperature
not the luminosity.
Whether or not it violates the Hayashi limit depends on whether 
the adopted temperature, inferred from the strength of the TiO bands or an 
atmospheric model, is indicative of the star's ill-defined surface or its
wind. With the above temperatures,  VY CMa is on the edge or just inside the 
Hayashi limit as plotted in Figure 1 in Massey et al., but the standard Hayashi 
limit applies to hydrostatic atmospheres.
Non-spherical outflows and a resulting dense wind as in VY CMa may affect the
Hayashi limit's location on the HR Diagram. 
In this regard, note that on this same figure, at their preferred luminosity for VY CMa, 
the star is also right on the edge of the Hayashi limit.  

Another  of Massey et al's problems with VY CMa is its large size. They point
out, however, that Monnier et al (2004) derived a radius of 3000 R$_{\odot}$ from 2 $\mu$m
interferometry.  Monnier himself, in the  Massey et al  paper, offered a couple 
of possible explanations for the apparent large size, including significant
structure in the dust shell  on the scale of the stellar diameter. This is probably 
correct (Humphreys et al 2005). Given  my
arguments above,  3000 R$_{\odot}$ is probably not the actual size of the imbedded star. 
 Adopting this radius with VY CMa's luminosity gives an ``effective'' temperature of 
$\sim$ 2700$\arcdeg$  which Massey et al say is too low. Alternatively,   
with the  apparent temperatures given above,  the radius is 1800 to 2100 R$_{\odot}$.  
In either case,  VY CMa is obviously very luminous, cool and big.  

\section{Concluding Remarks}

With its extraordinary high mass loss rate ( 2 -- 3  x $10^{-4}$ $M_{\odot}$ yr$^{-1}$, 
Danchi et al 1994), extensive circumstellar ejecta and 
discrete ejection episodes over the past 1000 yrs (Smith et al 2001, 
Humphreys et al 2005), VY CMa is undoubtedly not in hydrostatic equilibrium. Its extreme 
characteristics, strong wind and evidence for surface activity are more easily appreciated
at its position near the upper luminosity boundary  in the HR Diagram and near the Hayashi 
limit, than as a less luminous red supergiant ($\sim$ 15M$\sun$) as 
Massey et al suggested.

VY CMa is not alone.  In previous papers  (Humphreys et al 1997, Smith et al 2001) I  have  
pointed out that about 10 known M-type supergiants in the Milky Way and Local Group 
galaxies have M$_{bol}$ brighter than -9 mag. About half of these stars, 
in the Milky Way and in the LMC, are known OH/IR stars with strong maser emission some
of which are optically obscured. 
We do not know if this is a short--lived high mass loss phase that all red supergiants pass
through similar to the optically obscured carbon and M stars near the top of the AGB, 
or if it occurs  only in the most massive ones, due perhaps to enhanced instability
near the upper luminosity boundary. Present observations of VY CMa do not let us
distinguish whether it is in the process of creating an
optically thick cocoon or expelling it in transition back to warmer temperatures. 
VY CMa is undoubtedly one of the most imprtant stars for understanding the high mass
loss episodes in the final stages of massive star evolution.

\clearpage

%Figure 1
\begin{figure}
\epsscale{1.0}
\plotone{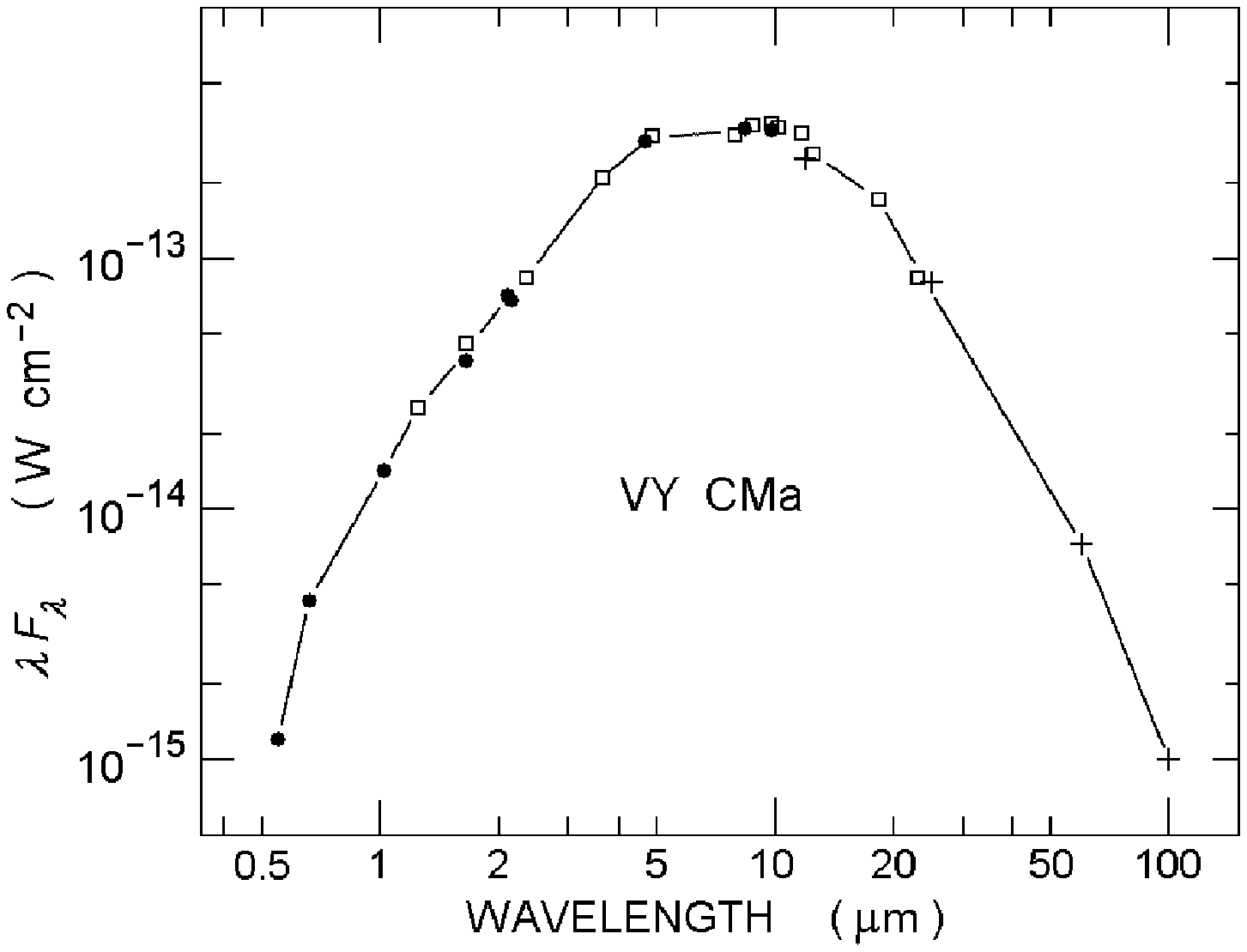}
\caption {The spectral energy distribution of VY CMa. The solid circles are fluxes 
measured from HST optical images and near-IR groundbased images (Table 3 in Smith et 
al (20010), the open squares are standard broadband IR photometry (Table 4 in Smith et al (2001), and the +'s are the IRAS 12 to 100$\mu$m fluxes.}
\label{Spectrum}
\end{figure}

%% The following command ends your manuscript. LaTeX will ignore any text


\begin{thebibliography}{}
\bibitem[Danchi et al. (1994)]{Dan94}Danchi, W.C., Bester, M., Degiacomi, C.G., Greenhill, L.J., \& Townes, C.H.  1994, \aj, 107, 1469
\bibitem[Flower 1977]{PF77}Flower, P. 1977, \aap, 54,32
\bibitem[Herbig 1970a]{GH70a}Herbig, G.H. 1970a, Mem. Soc. Roy. Liege, 19, 13
\bibitem[Herbig 1972]{GH72}Herbig, G.H.  1972, \apj, 172, 375
\bibitem[Humphreys 1975]{RMH75}Humphreys, R. M. 1975, \pasp, 87, 433
\bibitem[Humphreys \& Davidson (1979)]{HD79}Humphreys, R.M. \& Davidson, K. 1979, \apj, 232, 409
\bibitem[Humphreys \& Davidson (1994)]{HD94}Humphreys, R.M. \& Davidson, K. 1994, \pasp, 106, 1025
\bibitem[Humphreys, R. M. et al 2005]{RMH05}Humphreys, R. M., Davidson, K., Ruch, R. \& 
Wallerstein, G. 2005, \aj, 129, 492 
\bibitem[Humphreys \& McElroy 1984]{HM84}Humphreys, R. M. \& McElroy, D. B. 1984, \apj, 284, 565
\bibitem[Humphreys et al 1997]{RMH97}Humphreys, R.M., et al. 1997, \aj, 114, 2778 
\bibitem[Kwok 1976]{Kwok76}Kwok, S. 1976, JRASC, 70, 49 
\bibitem[Lada \& Reid 1978]{LR78}Lada, C.J. \& Reid, M.J.  1978, \apj, 219, 95
\bibitem[Le Sidaner \& Le Betre 1996]{LL96}Le Sidaner, P. \& Le Betre, T. 1996, \aap, 314, 896
\bibitem[Levesque et al 2005]{LM05}Levesque, E.M., Massey, P., Olsen, K. A. G., Plez, B., Josselin, E., Maeder, A., \& Meynet, G. 2005, \apj, 628, 973  
\bibitem[Marvel 1997]{Mar97}Marvel, K.B.  1997, \pasp, 109, 1286
\bibitem[Massey et al 2006]{PM06}Massey, P., Levesque, E. M., \& Plez, B. 2006, \apj, 646, 1203
\bibitem[Monnier et al 2004]{Mon04}Monnier, J. D. et al. 2004, \apj, 605, 436 
\bibitem[Schuster et al 2006]{Sch06}Schuster, M. T., Humphreys, R. M., \& Marengo, M 2006, \aj, 131, 603 
\bibitem[Smith et al 2001]{Smi01}Smith, N., Humphreys, R. M., Davidson, K., Gehrz, R.  D., Schuster, M. T. \& Krautter, J.  2001, \aj, 121, 1111
\bibitem[Van Blerkom \& Van Blerkom 1978]{VB78}Van Blerkom, J. \& Van Blerkom, D. 1978, \apj, 225, 482
\bibitem[Wallerstein 1977]{GW77}Wallerstein, G. 1977, \apj, 211, 170


\end{thebibliography}
\end{document}